\documentclass[amsmath,amssymb,aps,prb,reprint,superscriptaddress,onecolumn]{revtex4-1}

\usepackage{graphicx}
\usepackage{dcolumn}
\usepackage{bm}
\usepackage{subfigure}
\usepackage{url}
\usepackage{float}
\usepackage{lipsum}
\usepackage{color}

\begin{document}

\preprint{APS/123-QED}

\title{Elliptical vortex and oblique vortex lattice in the FeSe superconductor based on the nematicity and mixed superconducting orders}

\author{Da-Chuan Lu} \thanks{dclu@smail.nju.edu.cn}
\affiliation{Nanjing University, Nanjing 210046, China}
\author{Yang-Yang Lv}
\affiliation{Nanjing University, Nanjing 210046, China}
\author{Jun Li} \thanks{junli@nju.edu.cn}
\affiliation{Nanjing University, Nanjing 210046, China}
\author{Bei-Yi Zhu}
\affiliation{Condensed Matter Physics, Institute of Physics, Chinese Academy of Sciences, Beijing 100190, China}
\author{Qiang-Hua Wang}
\affiliation{Nanjing University, Nanjing 210046, China}
\author{Hua-Bing Wang}
\affiliation{Nanjing University, Nanjing 210046, China}
\author{Pei-Heng Wu}
\affiliation{Nanjing University, Nanjing 210046, China}
\affiliation{Synergetic Innovation Center of Quantum Information and Quantum Physics, University of Science and Technology of China, Hefei 230026, China}

\date{\today}


\begin{abstract}
\noindent
The electronic nematic phase is characterized as an ordered state of matter with rotational symmetry breaking, and has been well studied in the quantum Hall system and the high-$T_c$ superconductors, regardless of cuprate or pnictide family. The nematic state in high-$T_c$ systems often relates to the structural transition or electronic instability in the normal phase. Nevertheless, the electronic states below the superconducting transition temperature is still an open question. With high-resolution scanning tunneling microscope measurements, direct observation of vortex core in FeSe thin films revealed the nematic superconducting state by Song \emph{et al}. Here, motivated by the experiment, we construct the extended Ginzburg-Landau free energy to describe the elliptical vortex, where a mixed \emph{s}-wave and \emph{d}-wave superconducting order is coupled to the nematic order. The nematic order induces the mixture of two superconducting orders and enhances the anisotropic interaction between the two superconducting orders, resulting in a symmetry breaking from $C_4$ to $C_2$. Consequently, the vortex cores are stretched into an elliptical shape. In the equilibrium state, the elliptical vortices assemble a lozenge-like vortex lattice, being well consistent with experimental results.
\end{abstract}



\maketitle
%
%
\thispagestyle{empty}

\section*{INTRODUCTION}
\noindent

In the newly discovered high superconducting transition temperature ($T_c$) iron-based family, the FeSe superconductors possess the simplest crystalline structure but attract much attention owing to multifarious physical properties \cite{kamihara2008iron,Haindl,Hsu,Paglione}. The $T_c$ of bulk FeSe crystal is as low as 8 K, while it can be considerably enhanced to above 37 K under high-pressure \cite{Medvedev}, electric field gating \cite{Shiogai}, or insetting the intercalation layer \cite{dong2014phase}. Particularly, the monolayer FeSe on SrTiO$_3$ was observed a dramatically high $T_c$ above the liquid point of nitrogen \cite{100Kge2015superconductivity}, which offers the possibility of breaking the record as those of cuprate family. The origin for the enhancement of $T_c$ is still an open question, while a common consensus has been proposed as an accompaniment to the modification of the Fermi surface. Therefore, studying on the electronic state of the FeSe system provides a perfect arena to understand the high-$T_c$ mechanism.\par

Different from the conventional superconductors, competing electronic orders such as unidirectional charge density wave and nematic order exist in both cuprate and iron-based superconductors. Among these, the nematic electronic order demonstrates a spontaneous symmetry breaking from $C_4$ to $C_2$ symmetry (the order parameters remain invariant under the inversion, the $D_{4h}$ group can be viewed as $C_4$), which has been generally considered as a strong correlation with the fundamental unsolved electronic issue in Fe-based superconductors, especially in recent work on the FeSe system \cite{mcqueen2009tetragonal,watson2015emergence}. For the FeSe bulk crystals, the structural transition from tetragonal to orthorhombic occurs at $T_s=90$ K, while the anisotropy of the electronic structure is not a consequence of the lattice distortion, but a result of the microscopic mechanism such as spin fluctuation or orbital ordering. Researches on the nematic order in iron-based superconductors have generally supported the spin-fluctuation origin. However, because of the absence of long-range magnetic order in the FeSe system, orbital ordering is probably the origin for the electronic transition. Moreover, recent angle-resolved photoemission spectroscopy (ARPES) results showed the emergence of the nonequivalent energy shifts of $xz/yz$ orbital bands below $T_s$ \cite{shimojima2014lifting,watson2015emergence}, implying the orbital origin of the structural transition.\par

Furthermore, similar to the nematic order, the superconducting pairing symmetry strongly relates to detailed electron-electron interaction. To be specific, the orbital order with inter-orbital electron-electron interactions would favor a sign-preserving $s$-wave pairing, while spin fluctuation with intra-orbital interaction for a sign-changing $s_{\pm}$-wave or $d$-wave. Recent ARPES and Scanning Tunneling Microscopy (STM) results suggested the sign-changing pairing symmetry such as $s_{\pm}$- wave or $d$-wave in FeSe, implying that the magnetic fluctuations may still assist the superconducting pairing \cite{nakayama2014reconstruction,moore2015evolution}. With the high-resolution STM measurement,  the elliptical vortices have been directly observed at superconducting state on the FeSe bulk samples \cite{song2011direct}, for which the extremely weak structure distortion ($\sim$ 0.5\%) can hardly induce such pronounced anisotropy, while the nematic order and superconducting order parameters are expected to play the important roles. Ginzburg-Landau (GL) theory firstly can offer a phenomenological way to investigate the vortices in superconductors with the $s$-wave symmetry. The GL theory itself can be derived exactly from the microscopic BCS theory \cite{gor1959microscopic_derive}. By means of the Gorkov's derivation and symmetry analysis, the GL theory has been generalized into several pairing symmetries such as $s+id$ \cite{sid_berlinsky1995ginzburg, sid_feder1997microscopic, sid_ren1995ginzburg, sid_ren1996ginzburg, sid_xu1995ginzburg}, $p$-wave \cite{pwave_heeb1999ginzburg}, and so on. Among these symmetry models the $s+id$, where the extended $s_{\pm}$-wave competes with the $d$-wave pairing order, is generally used to investigate the iron-based superconductors\cite{lee2009pairing,chowdhury2011nematic, competing_orderkivelson2002competing}.

In this work, we construct the GL type free energy which contains the nematic order, $s$-wave and $d$-wave superconducting orders with up to $4^{th}$ order interactions. The time-dependent GL (TDGL) equation is derived from the free energy to describe the FeSe system. By implementing the open boundary condition, our simulation reveals the configuration and dynamics of the elliptical vortex and the nematic order. With the periodical boundary condition, the oblique vortex lattice rather than a triangular one is found. Our simulation results have a good agreement with the previous experiment in the configuration of the single vortex and the vortex lattice \cite{song2011direct}. The presence of the nematic order can break the symmetry from $C_4$ to $C_2$ and enhance the superconductivity. The symmetry allowed trilinear term will enhance the anisotropy of the superconducting order and induce the nearly degeneracy of $s$-wave and $d$-wave \cite{livanas2015nematicity,fernandes2013nematicity}.  \par

\section*{RESULTS}\label{section4_result}
\noindent High-resolution STM and scanning tunneling spectroscopy (STS) experiments provide the possibility for further investigation on the single vortex, for which the vortices configuration can be reconstructed as well studied in various superconductors \cite{hess1989scanning,karapetrov2005direct,hess1990vortex,yin2009scanning,de1997scanning,shan2010observation}. In the previous work by Song $et$ $al$. \cite{song2011direct}, the vortices and vortex lattice in the FeSe superconductors were directly observed, and the vortex core was found in an elliptical shape, where the stretched direction is along one of the Fe-Fe bonds.\par

With the open boundary condition, the interplay between the anisotropic vortices and finite geometric region are investigated in the present work. Although one can hardly observe the evolution and dynamics of the vortices and nematic order in realistic experiments, the real-time simulation results can provide an approach to understand the motion of the vortices. By solving the TDGL equations, the results show that nematic order breaks the symmetry from $C_4$ to $C_2$ during the evolution.\par

By using the periodic boundary condition, the vortex lattice is also investigated. Based on the simulation, the vortices favor an oblique lattice rather than the triangular lattice; this is due to the trade-off between the twofold symmetry of the repulsive interaction and the closet packing. The simulation results are consistent with the experimental data.\par

\subsection*{FINITE REGION AND VORTEX CONFIGURATION}\label{section7_ourmodel_result}
\noindent

Previous works suggest that the pairing symmetry is probably $s_{\pm}$ wave or $d$ wave, they both can be described by the addition of the isotropic and the anisotropic superconducting order. The isotropic order parameter is coupled to the anisotropic order parameter by the interaction \cite{sid_ren1995ginzburg,sid_berlinsky1995ginzburg,joynt1990upward},
\begin{equation}\label{interaction}
{F_{{\mathop{\rm int}} }} = \frac{\gamma }{2}\left( {{\Pi _x}{\psi _s}\Pi _x^*\psi _d^* - {\Pi _y}{\psi _s}\Pi _y^*\psi _d^* + c.c.} \right)
\end{equation}
where the two complex fields ${\psi _s}, {\psi _d}$ stand for the \emph{s}-wave component and \emph{d}-wave component in the mixed superconducting order, and $\gamma$ is the coupling constant. ${\Pi } = \left( { - i\hbar \nabla  - e^*{\bf{A}}} \right)$ is the gauge invariant derivative, $\nabla$ is the del operator, $\mathbf{A}$ is the magnetic vector potential, and $e^*$=2$e$ is the charge of the superconducting charge-carriers, where $e$ is the electron charge. This term is invariant under rotation of $\pi/2$, when taking the integration by parts,

\begin{equation}\label{interaction_part_integral}
{\psi _s}\left( {\Pi _x^2 - \Pi _y^2} \right)\psi _d^* + c.c.
\end{equation}
and thus,

\begin{equation}\label{interaction_part_integral2}
\begin{array}{l}
{\psi _s}\left( {\Pi _x^2 - \Pi _y^2} \right)\psi _d^* + c.c. \to \\
{\psi _s}\left[ { - \left( {\Pi _x^2 - \Pi _y^2} \right)} \right]\left( { - \psi _d^*} \right) + c.c. \to {\psi _s}\left( {\Pi _x^2 - \Pi _y^2} \right)\psi _d^*+ c.c.
\end{array}
\end{equation}

The above type of anisotropic interaction is used in our model. Besides the mixed superconducting order, a real field $\phi$ stands for the nematicity order, which competes with the mixed superconducting order. Because the higher order terms are negligible, the free energy which is up to 4$^{th}$ order can be described as \par
\begin{equation}\label{freeenergy}
\begin{array}{l}
f = {f_s} + {f_d} + {f_\phi } + {f_{{\rm{int}}}}\\
{f_{s,d}} =  - {\alpha _{s,d}}{\left| {{\psi _{s,d}}} \right|^2} + \frac{{{\beta _{s,d}}}}{2}{\left| {{\psi _{s,d}}} \right|^4} + \frac{1}{{2{m_{s,d}^*}}}{\left| {{\Pi}{\psi _{s,d}}} \right|^2}\\
{f_\phi } =  - {\alpha _\phi }{\phi ^2} + \frac{{{\beta _\phi }}}{2}{\phi ^4} + \frac{1}{{2{m_\phi }}}{\left| {\nabla \phi } \right|^2}\\
{f_{{\rm{int}}}} = {\gamma _1}{\left| {{\psi _s}} \right|^2}{\left| {{\psi _d}} \right|^2} + {\gamma _2}\left( {\psi _s^{*2}\psi _d^2 + c.c.} \right)\\
 + {\gamma _3}\left( {{\Pi _x}{\psi _s}\Pi _x^*\psi _d^* - {\Pi _y}{\psi _s}\Pi _y^*\psi _d^* + c.c.} \right)\\
 + {\lambda _1}\phi \left( {\psi _s^*{\psi _d} + c.c.} \right) + {\lambda _2}{\phi ^2}{\left| {{\psi _s}} \right|^2} + {\lambda _3}{\phi ^2}{\left| {{\psi _d}} \right|^2}
\end{array}
\end{equation}
where ${\alpha _i}$, ${\beta _i}$ are the parameters describing the Landau phase transition and $i = s$, $d$ and $\phi $. Considering ${T_s} < {T_d} < {T_\phi }$, the ${\alpha _i}$ submits to
${\alpha _s} < {\alpha _d} < {\alpha _\phi }$. ${\gamma _j}$ ($j$ = 1, 2, 3) is the coupling constant between $s$-wave and $d$-wave components, and $\lambda_k$ ($k$= 1-3) is the coupling constant between the superconducting order parameters and the nematic order. $m_i^*$ ($i=s,d$) is the mass of the superconducting charge-carriers, and the microscopic electron pairing theory of superconductivity implies that $m_i^*$=2 $m_i$, where $m_i$ is the electron mass. $m_\phi$ represents the effective mass of the nematic order.\par

Instead of directly making the isotropic order parameter coupled to the nematic order, our model (Eq.~\ref{freeenergy}) suggests that the nematic order triggers off the mixture of $s$-wave and $d$-wave components, and the anisotropic interaction between \emph{s}-wave and \emph{d}-wave components causes larger anisotropy. Meanwhile, the trilinear term ${\lambda _1}\phi \left( {\psi _s^*{\psi _d} + c.c.} \right)$ will enhance the anisotropy. Besides, according to the previous mean-field analysis, this term may induce nearly degeneracy of $s$-wave and $d$-wave, which was also supported by the spin-fluctuation model \cite{livanas2015nematicity,fernandes2013nematicity}. This is different from the \emph{p}-type 122-system, where the degeneracy of $s$-wave and $d$-wave is due to the close critical temperatures \cite{li2017nematic}. \par\par

In the following simulation, the phase transition parameters in the Eq.~\ref{freeenergy} are set to be $\alpha_s$ = 1.0, $\alpha_d$ = 1.5, $\alpha_\phi$ = 2.0, $\beta_i$ = 1, $i$ = $s$, $d$, and $\phi$, which are based on the superconductivity and nematicity transition temperature \cite{nakayama2014reconstruction,huynh2014electric,mcqueen2009tetragonal}.  The coupling constant of the anisotropic interaction is set as $\gamma_3$ = 0.2, while other coupling constants are set to be 0.4 based on the consideration of convergency. $\lambda_2,\lambda_3$ are negative. The effective masses are set to be $m_s=1, m_d=2, m_\phi=4$.\par

The initial condition is quite important for convergency of the non-linear partial differential equations, though different initial conditions will arrive at the same stable state in this type partial differential equations. The initial states of the complex order parameters are set to be proportional to ${\psi _0} = \frac{{x - {x_0} + i\left( {y - {y_0}} \right)}}{{\sqrt {{{\left( {x - {x_0}} \right)}^2} + {{\left( {y - {y_0}} \right)}^2}} }}$. For the $s$-wave component, the initial state is $\psi_0$, while it is relatively small for the $d$-wave component. The initial state of the nematic order is set to be $0.5$, and all the vector potentials are set to be $0$.\par

\subsubsection*{EVOLUTION OF THE VORTICES AND THE NEMATIC ORDER}\label{section8_timedependent_sol}
\noindent The vortex core can be visualized by atomically resolved STM measurements. Moreover, by applying the specific periodical magnetic field, it is possible to view the motion of the vortex \cite{timmermans2014dynamic}. However, it is still difficult to conduct real-time observation on the evolution and dynamics of the vortices. Time-dependent simulation can be applied to simulate how the vortices generate from the boundary and how they move and interact with each other. Through the real-time simulation, deep understanding of the vortex dynamics can be achieved and further applications can be simulated, such as modification of the sample to enhance the critical current.\par

Different from the other type-II superconductors, the elliptical vortices in FeSe sample is related to the $C_4\rightarrow C_2$ symmetry breaking, where the nematic order plays an important role on enhancing the symmetry breaking in the superconducting state \cite{wang2015nematicity}. With the real-time simulation, the transition from $C_4$ symmetry to $C_2$ symmetry is revealed (see Supplementary videos). \par

The real-time evolution of the elliptical vortices is shown in Fig. 1. At the beginning, the pattern of the $s$-wave component shown in Fig. 1a.1 is quite similar to that in typical type-II superconductors \cite{alstrom2011magnetic,cabral2004vortex,berdiyorov2006vortex,schweigert1998vortex}, the magnetic field penetrates into the sample from the edge. Due to the Bean-Livingston barrier \cite{burlachkov1993magnetic,deo1999hysteresis}, the vortices cannot immediately get into the sample. The interaction of intra-vortices is qualitatively repulsive and the vortices could only locate along the edges.\par

The enhanced magnetic field will force the vortices to penetrate into the sample. As a result, Fig. 1a.2 demonstrates the arrangement of four vortices. The system will gradually achieve its minimum free energy by rearranging the vortices. However, with the nematic order competing with the mixed superconducting order, the situation is quite different. Fig. 1a.3-6 depict the intermediate stage, where the $C_4\rightarrow C_2$ symmetry breaking happens, because the nematic order mixes the two superconducting order and enhances the anisotropic interaction. Meanwhile, the interaction terms $\lambda_1$ and $\gamma_{1,2}$ compete with each other, $\lambda_1$ favors a large separation between $s$-wave and $d$-wave, while the other one favors small separation. Fig. 1a.7 demonstrates the equilibrium state of this system, the two vortices are both elliptical and repulse with each other in short range.\par

Different from the $s$-wave component, the initial state of the $d$-wave component is relatively small and then induced to be anisotropic by the nematic order. The vortices of $d$-wave component are slightly less eccentric due to its fourfold tendency. Apart from the superconducting order, the nematic order shown in Fig. 1c.1-7 exhibits strong $C_2$ symmetry at first. It then has fourfold symmetry due to the interaction with $s$-wave component. The final state shown in Fig. 1c.7 has $C_2$ symmetry and is less eccentric than the superconducting order.\par

\subsection*{VORTEX LATTICE}\label{section_vortex_lattice}

Vortices arrange into the vortex lattice in an infinite region. Though it is impossible to simulate the complicated TDGL equations in an infinite region, simulation on the single unit cell with periodic boundary condition can make the investigation of the vortex lattice possible \cite{xu1996structures,wang1996simulating,hong2004simulating}. By extending the unit cell according to the periodicity, the vortex lattice can be recovered.\par
Defining the ratio of the side lengths of the rectangular unit cell as $r$, $r=\sqrt{3}$ corresponds to triangular lattice which is typical for most type-II superconductors, as shown in Fig. 2a. Early study reported that the oblique vortex lattice \cite{xu1996structures,sid_berlinsky1995ginzburg,franz1996vortex,wang1996simulating}, $r<\sqrt{3}$, for $s+id$ model costs less energy than the triangular one which is ascribed to the fourfold symmetry of the system. To be specific, the system tends to preserve the fourfold symmetry while the closet packing between the vortices leads the vortex lattice to be triangular. As a result, the vortex lattice favors the oblique one by making the trade-off between preserving the fourfold symmetry and the triangular lattice. While in our case, the nematic order breaks the $C_4$ symmetry to $C_2$, and thus the system finds the balance between the twofold symmetry and the triangular lattice. Therefore, it favors the lattice with $r>\sqrt{3}$. \par

In the simulation, the normalized magnetic field is set as $4\pi$, which allows two vortices in the one unit cell. The area of the unit cell is set to be $16 \lambda^2$. By varying the ratio $r$, the minimum energy density $f=F/L_x/L_y$ is achieved at $r=2.81$, as shown in Fig. 2. It is hard to realize the real-time detection in experiments and compare the vortex dynamics with the simulation, but the equilibrium state where the vortices form a stable vortex lattice can be compared with the simulation results. Based on the parameters provided above, the simulated oblique vortex lattice $r = 2.81$ is in agreement with the previous result $r \sim 2.80$ in the experiment where many vortices are observed in FeSe under applied magnetic field of 8 T \cite{song2011direct}.\par

\section*{DISCUSSION}
\subsection*{ANALYTICAL TREATMENT OF THE ANISOTROPY} \label{section10_analytic}
\noindent According to the previous simulation results, the anisotropy of the interaction serves to the elliptical shape of the vortices. Since the London penetration depth is considerably large than the coherence length $\lambda\gg \xi$, the coupling to the electromagnetic field can be neglected. Given that the variation of the nematic order is small, and the vortices experience a uniform nematic order away from the origin, thus nematic order is set to be a stationary field, $\phi=\phi_0$, and satisfy $\phi_0\rightarrow -\phi_0$ under the rotation of $\pi/2$. The equations of the superconducting order are,
\begin{equation}\label{approx_equ}
\begin{array}{l}
\left( {{\lambda _2}\phi _0^2 - {\alpha _s}} \right){\psi _s} + \beta {\left| {{\psi _s}} \right|^2}{\psi _s} + {\left| {{\psi _d}} \right|^2}\left( {{\gamma _1}{\psi _s} + 2{\gamma _2}\psi _s^*} \right)
 + {\gamma _s}\left( {\partial _x^2 + \partial _y^2} \right){\psi _s} + {\gamma _3}\left( {\partial _x^2 - \partial _y^2} \right){\psi _d} + {\lambda _1}{\phi _0}{\psi _d} = 0\\
\left( {{\lambda _3}\phi _0^2 - {\alpha _d}} \right){\psi _d} + \beta {\left| {{\psi _d}} \right|^2}{\psi _d} + {\left| {{\psi _s}} \right|^2}\left( {{\gamma _1}{\psi _d} + 2{\gamma _2}\psi _d^*} \right)
 + {\gamma _d}\left( {\partial _x^2 + \partial _y^2} \right){\psi _d} + {\gamma _3}\left( {\partial _x^2 - \partial _y^2} \right){\psi _s} + {\lambda _1}{\phi _0}{\psi _s} = 0
\end{array}
\end{equation}
Where $\gamma_{s,d}=\frac{\hbar^2}{2 m_{s,d}}$. A direct observation on Eq.~\ref{approx_equ} reveals that $\lambda _2,\lambda _3$ and $\phi_0^2$ will change the critical point of the phase transition in this system, as the minimum point of the potential is at ${\psi _{s,\min }} =  \pm \sqrt {\left( {{\alpha_s} - \frac{{{\lambda _2}}}{2}\phi _0^2} \right)/{\beta_s}}$, ${\psi _{d,\min }} =  \pm \sqrt {\left( {{\alpha_d} - \frac{{{\lambda _3}}}{2}\phi _0^2} \right)/{\beta_d}} $. The coefficients of the term $\psi_s$ and $\psi_d$ are non-zero and different for most situations. Because $\lambda_2,\lambda_3$ are negative, the presence of the nematic order will enhance the superconductivity \cite{Li2016}.\par

The interesting question is that the $\lambda_1$ term will break the rotation symmetry. When rotating $\pi/2$, the $\lambda_1$ term picks up a different sign compared with other terms. Such term enhances the anisotropy of the vortices.\par

Polynomial terms in Eq.~\ref{approx_equ} do not contribute to the anisotropy but the gradient terms and the nematic order will. Without losing universality, we set $\gamma_s = \gamma_d = \gamma$. Eq.~\ref{approx_equ} can be reformulated as,
\begin{equation}\label{approx_equ_reformed}
\left( {\gamma \left( {\partial _x^2 + \partial _y^2} \right) + {\gamma _3}\left( {\partial _x^2 - \partial _y^2} \right)} \right)\psi  + {\lambda _1}{\phi _0}\psi  + P\left( {{\psi _s},{\psi _d}} \right) = 0
\end{equation}
where $\psi=\psi_s+\psi_d$, $P$ is the polynomial of $\psi_s$ and $\psi_d$. Ignoring the nematic order $\phi_0$, the solution is actually a elliptical vortex, which can be seen by transforming the gradient terms to a Laplacian under the coordinate transformation,
\begin{equation}\label{coordinate_transformation}
x' \to \frac{1}{{\sqrt {1 + \gamma_3 /\gamma } }}x,y' \to \frac{1}{{\sqrt {1 - \gamma_3 /\gamma } }}y
\end{equation}
Due to Eq.~\ref{coordinate_transformation}, the coordinate is elongated along $y$ direction and contracted along $x$ direction as shown in Fig.~3. Therefore, the symmetry is broken into $C_2$ once returning to the original coordinate.\par

However, the nematic order obeys $\phi_0 \rightarrow -\phi_0 $ under rotation of $\pi/2$, it is impossible to view $x$-direction and $y$-direction equivalently, because the nematic order offers a angle dependent term. To capture the feature, $\phi_0$ is set to be $k(x^2-y^2)$, where $k$ is a constant. Fig.~4 shows that turning on the nematic order makes the elliptical vortex more anisotropic.

\subsection*{COMPARISON WITH OTHER SYSTEMS} \label{section11_discuss}
The anisotropic electronic structure is normal in some high-$T_c$ superconductors. For instance, in the cuprate family, YBa$_2$Cu$_3$O$_{7-\delta}$ (YBCO) has a tetragonal to orthorhombic phase transition at the under-doped level of oxygen, resulting in a symmetry breaking from $D_{4h}$ to $D_{2h}$ along the Cu-O chains \cite{lei1993elastic}. Here, the structure transition temperature is considerably higher than the superconducting transition temperature $T_c$. The previous works on the YBCO have constructed the GL free energy which obeys $D_{2h}$ symmetry and found the elliptical vortex along the $b$-axis\cite{heeb1996vortices,han1999vortex}. In the present FeSe system, orthorhombic phase happens after a structural phase transition at 90 K. However, it is argued that the small crystalline distortion ($\sim 0.5\%$) itself cannot lead to such large anisotropy in electronic structure \cite{song2011direct,fernandes2014drives}. Therefore, the nematic order may originate from the strongly interacting fermion system, where the small symmetry breaking term will be promoted to finite magnitude when lowering the energy scale, say, temperature. Thus, different from the explicit symmetry breaking treatment in YBCO, the symmetry breaking nematic order in FeSe is temperature dependent and should be treated as a competing order which competes with the $D_{4h}$ superconducting order \cite{competing_orderkivelson2002competing}. On the other hand, the elliptical vortex core is along the Fe-Fe bonds, being considerably different from that along the $b-$axis in YBCO.

In some iron-based superconductors, the superconducting nematic transition temperature was also found below the superconducting transition temperature, such as \emph{p}-type 122-system iron pnictide superconductors\cite{li2017nematic}. The nematic superconducting state in this superconductor is induced by the small symmetry breaking term in the normal phase which qualitatively differs from the ordinary nematicity observed in the orthorhombic structural phase. The superconducting state nematicity occurs just after the onset of superconducting transition and reveals an anisotropy shifted by $\pi/4$ from those of the normal nematic state \cite{li2017nematic}. Meanwhile, the superconducting transition temperature resembles a low energy scale, hereby, the lower nematic transition temperature may correspond to an infrared phenomenon of the strongly interacting fermion system.\par
However, one can hardly identify which order occurs first. When the superconducting transition temperature $T_c$ is lower than the nematic transition temperature $T_n$, it can be expected that the nematic order still exists in superconducting state and strongly interacts with the superconducting orders, which are shown in Fig.~5a. For the case of $T_c >T_n$, there is an intermediate region $T_c>T>T_n$, where the nematic order is absent in the superconducting state, and the profile of the vortex core is shown in Fig.~5b. Without the nematic order, the mixture of $s$-wave and $d$-wave makes the vortex slightly anisotropic. However, with the nematic order occurring below $T_n$, the anisotropy of the superconducting vortex is enhanced. In the simulation, the high order correction term is added into the free energy.

\section*{METHODS}\label{section1_model}
\noindent We begin by constructing the GL type free energy. Intuitively, the nematic order competes with the mixed superconducting order parameters, where isotropic \emph{s}-wave and anisotropic \emph{d}-wave components are considered. We mainly focused on the 2-dimensional (2D) geometry due to the quasi-2D feature for the Fe-based superconductors, and the 2D GL free energy can mostly capture the ingredients in the high-$T_c$ superconductors. By taking the variation of the free energy, the TDGL equations are derived. The TDGL was constructed to investigate the dynamics of the vortices in the dirty limit where the penetration depth $\lambda$ is greatly larger than the coherence length $\xi$ \cite{schmid1966time}. For the FeSe case, due to $\lambda \sim 500$ nm and $\xi \sim 5$ nm \cite{terashima2014anomalous, bendele2010anisotropic ,khasanov2008evidence}, the dirty limit construction of TDGL is valid. Numerical results of the TDGL equations with the open boundary condition, given by the finite element method, can reveal the shape, configuration and dynamic properties of the vortices. The numerical solutions with periodic boundary condition provide the vortex lattice in the equilibrium state. Our theoretical calculation results show the elliptical vortices and oblique vortex lattice, which are in agreement with the experiment \cite{song2011direct}. In the following formalism, the $a-$ and $b-$axis, or $x-$ and $y-$axis are defined along either of the Fe-Fe bond directions as shown in Fig. 6, for which the directions of the Fe-Fe bonds keep the symmetry of the nematicity.

\subsection*{FREE ENERGY}\label{section2_free_energy}
\noindent Competing order such as nematicity can strongly interact with the superconducting order parameters. Chowdhury $et$ $al.$ investigated the anisotropic interplay between the competing order and the single isotropic superconducting order \cite{chowdhury2011nematic}, and they argued that the different effective masses which are induced by the anisotropic interaction could lead to the anisotropic vortex. However, the pairing symmetry in FeSe is sign-changing $s_{\pm}$-wave or $d$-wave suggested by the recent experiments \cite{nakayama2014reconstruction,moore2015evolution}. Meanwhile, the $s+id$ model could suitably describe the iron-based superconductors \cite{lee2009pairing}. Here, the mixed superconducting order together with the anisotropic interaction can also cause an anisotropic vortex core, where the isotropic \emph{s}-wave order parameter interacts with the anisotropic \emph{d}-wave order parameter. The existence of nematic order will mixed the two superconducting order and significantly enhance the anisotropy. Thus, the nematic order can enhance the small anisotropic interaction between the superconducting orders to form an extremely elliptical vortex.

Previous work investigated the special trilinear term thoroughly. It turns out if the coupling constant $\lambda_1=0$ or very small, the system may favor $s+id$ symmetry, but for large nematic fluctuation, the intermediate state has $s+d$ symmetry\cite{fernandes2013nematicity} character. The free energy (Eq.~\ref{freeenergy}), including both self-energy and interaction energy, remains invariant under the mirror reflection and the rotation of $\pi /2$.
\begin{equation}\label{symmetry}
x \to y,y \to  - x,{\psi _s} \to {\psi _s},{\psi _d} \to  - {\psi _d}, \phi \to -\phi
\end{equation}
However, the ${\gamma _3}$ term in the interaction causes different effective masses along the two directions, and consequently, results in the anisotropic vortex cores and affects the arrangement of the vortices, namely, the vortex lattice. Based on the consideration of symmetry, up to $4^{th}$ order, it is impossible to turn on the direct interaction between the nematic order and the anisotropic gradient term. Nevertheless, the nematic order can tune the stationary part of the superconducting orders in the free energy and let them mixed, thus the anisotropic interaction ${\gamma _3}$ enhanced. The $\lambda_1$ term will also enhance the anisotropy as explained in Section Analytical Treatment of the Anisotropy. Because the $\lambda_2, \lambda_3$ are negative, the presence of the nematic order will enhance the superconductivity, such as increasing the transition temperature.

\subsection*{TIME-DEPENDENT GINZBURG-LANDAU EQUATION}\label{section3_method}
\noindent The TDGL equation can be obtained by taking the variation of the free energy as follow,
\begin{equation}\label{time_dependent_ginzburg_landau_equ}
\begin{array}{l}
\frac{{{\hbar ^2}}}{{2{m_i}{D_i}}}\left( {\frac{\partial }{{\partial t}} + i\frac{e}{\hbar }\Phi } \right){\psi _i} =  - \frac{{\delta f}}{{\delta \psi _i^*}}\\
\frac{{{\hbar ^2}}}{{2{m_\phi }{D_\phi }}}\frac{{\partial \phi }}{{\partial t}} =  - \frac{{\delta f}}{{\delta \phi }}\\
\sigma \left( {\frac{{\partial {\bf{A}}}}{{\partial t}} + \nabla \Phi } \right) =  - \frac{{\delta f}}{{\delta {\bf{A}}}} - \frac{1}{{4\pi }}\nabla  \times \nabla  \times {\bf{A}}
\end{array}
\end{equation}
where ${D_i}$ ($i$ = $s$, $d$, $\phi$) is the phenomenological diffusion coefficients, and $\Phi$ is the scalar potential of the electromagnetic field. \par

In the open boundary conditions, the superconductivity-vacuum boundary can be obtained directly from the variation of the free energy as
\begin{equation}\label{boundaryconditions}
\begin{array}{l}
\left( {\frac{\hbar }{i}\left( {{l_{x,i}}\frac{\partial }{{\partial x}}{\bf{\hat x}} + {l_{y,i}}\frac{\partial }{{\partial y}}{\bf{\hat y}}} \right) - q{\bf{A}}} \right){\psi _i} \cdot {\bf{n}} = 0\\
\nabla \phi  \cdot {\bf{n}} = 0\\
\nabla  \times {\bf{A}} = {{\bf{B}}_{\bf{a}}}\\
\left( {\frac{{\partial {\bf{A}}}}{{\partial t}} + \nabla \Phi } \right) \cdot {\bf{n}} = 0,
\end{array}
\end{equation}
where ${\bf{\hat x}},{\bf{\hat y}}$ are the unit vector along $x$ and $y$ directions, respectively, and ${l_{x,i}} = 1 + \frac{{{m_i}}}{{{m_{sd}}}}$ and ${l_{y,i}} = 1 - \frac{{{m_i}}}{{{m_{sd}}}}$ correspond to the different effective mass along $x$ and $y$ directions, respectively. Thus, the dynamics property of the superconducting order is different along the $x$ and $y$ axis. \par

To solve the TDGL equations numerically, the complicated TDGL equations are normalized by introducing,
\begin{equation}\label{normalization}
\begin{array}{l}
\left( {x,y,z,t} \right) \to \left( {\lambda x',\lambda y',\lambda z',\frac{{{\xi ^2}}}{D}t'} \right),{\bf{A}} = \frac{\hbar }{{e\xi }}{\bf{A}}'\\
{\psi _i} = \sqrt {\frac{{{\alpha _i}}}{{{\beta _i}}}} {\psi _i}'\left( {i = s,d} \right),\sigma  = \frac{1}{{{\mu _0}D{\kappa ^2}}}\sigma '
\end{array}
\end{equation}
where the new quantities are labeled by prime, the spatial and temporal coordinates are scaled according to the $\lambda $ and the $\xi  = \frac{\hbar }{{\sqrt {2{m_s}{\alpha _s}} }}$, the GL parameter is defined as $\kappa  = \lambda /\xi $ and $\kappa \gg 1$ for the FeSe system \cite{terashima2014anomalous, bendele2010anisotropic ,khasanov2008evidence}. The dimensionless form contains the gauge invariant derivative, ${\Pi } =  - \frac{i}{\kappa }\nabla  - {\bf{A}}$.\par

The TDGL is invariant under the gauge transformation. Given an arbitrary function $\chi \left( {x,y,z,t} \right)$, and introduce the gauge transformation as,
\begin{equation}\label{gaugetransformation}
{\tilde \psi _i} = \psi {e^{i\kappa \chi }},{\bf{\tilde A}} = {\bf{A}} + \nabla \chi ,\tilde \Phi  = \Phi  - \frac{{\partial \chi }}{{\partial t}}.
\end{equation}
Because of the extra degree of freedom, the gauge should be fixed to achieve the definite equations. For the sake of the simplicity, the scalar potential $\Phi$ can be eliminated by choosing the London gauge, let,
\begin{equation}\label{fixedgauge}
\frac{{\partial \chi }}{{\partial t}} = \Phi
\end{equation}
thus the TDGL equations are no longer dependent on the scalar potential $\Phi$, and the electric field is ${\bf{E}} =  - \frac{{\partial {\bf{A}}}}{{\partial t}}$.\par

The former procedure derives the TDGL equations and corresponding open boundary conditions based on the free energy. To implement the finite element method, the complex order parameters are decomposed into the real and imaginary part; the vector potentials are decomposed into $x, y$ components. For the consequences of the non-linear feature, the mesh of the region is adaptively refined to achieve high accuracy. Solving TDGL is minimizing the total energy of the system, and the stable state will be reached after hundreds to thousands of the normalized time.

\subsection*{PERIODIC BOUNDARY CONDITION}\label{section_qbc}

\noindent The open boundary condition describes the superconductor-vacuum boundary straightforward, therefore, it is useful to investigate the finite size solution. When coming to the infinite size, the periodic boundary condition should be introduced on each unit cell. The complex order parameter and the vector potential should be modified from one unit cell to another which can ensure the gauge invariance \cite{du1992analysis}. Two lattice vectors are used to characterize the lattice, namely, ${{\bf{t}}_{1}}$ and ${{\bf{t}}_{1}}$. The complex order parameters will pick up a phase while additional term should be added to the vector potential from one unit cell to another, namely,
\begin{equation}\label{quasi-periodic boundary}
\begin{array}{l}
{\psi _i}\left( {{\bf{x}} + {{\bf{t}}_k}} \right) = {\psi _i}\left( {\bf{x}} \right){e^{i\kappa {g_k}}}\\
{\bf{A}}\left( {{\bf{x}} + {{\bf{t}}_k}} \right) = {\bf{A}}\left( {\bf{x}} \right) + \nabla {g_k}
\end{array}
\end{equation}
where, ${g_k} =  - \frac{1}{2}\left( {{{\bf{t}}_k} \times B{{\bf{k}}_3}} \right) \cdot {\bf{x}}$, $k$ = 1 and 2, and $B$ is quantized by $B = \frac{{2\pi n}}{{\kappa \left| \Omega  \right|}}$, in which $\left| \Omega \right|$ is the area of the unit cell and $n$ is integer.\par
The explicit form for a rectangular unit cell is,
\begin{equation}\label{quasi-periodic boundary_explicit}
\begin{array}{l}
{\psi _i}\left( {{L_x},y} \right) = {\psi _i}\left( {0,y} \right){e^{\frac{{i\phi y}}{{2{L_y}}}}}\\
{A_x}\left( {{L_x},y} \right) = {A_x}\left( {0,y} \right)\\
{A_y}\left( {{L_x},y} \right) = {A_y}\left( {0,y} \right) + \frac{\phi }{{2\kappa {L_y}}}\\
{\psi _i}\left( {x,{L_y}} \right) = {\psi _i}\left( {x,0} \right){e^{\frac{{ - i\phi x}}{{2{L_x}}}}}\\
{A_x}\left( {x,{L_y}} \right) = {A_x}\left( {x,0} \right) - \frac{\phi }{{2\kappa {L_x}}}\\
{A_y}\left( {x,{L_y}} \right) = {A_y}\left( {x,0} \right)
\end{array}
\end{equation}
where, $i = s$ and $d$, $\phi = 2 n \pi$ is the reduced vortex flux, $L_x$ and $L_y$ characterize the size of the unit cell. The variation of the vector potential is neglected, due to $\kappa \gg 1$.

\section*{DATA AVAILABILITY}
The data that support the findings of this study are available from the corresponding author upon reasonable request.

\section*{ACKNOWLEDGEMENTS}

The work was supported by the National Natural Science Foundation of China (61727805, 61501220, 61771234, 6151101183, 11227904, 11234006, 61521001), Jiangsu Provincial Natural Science Fund (BK20150561), Opening Project of Wuhan National High Magnetic Field Center (2015KF19), and the Fundamental Research Funds for the Central Universities.

\section*{ADDITIONAL INFORMATION}
\textbf{Supplementary videos:} Supplementary videos are available at \emph{npj Quantum Materials} website. \par
\noindent \textbf{Competing interests:} The authors declare that they have no competing financial interests.

\section*{AUTHOR CONTRIBUTIONS}
J.L. and B.Y.Z. proposed the idea. D.C.L. proposed the model and Q.H.W. improved the model. D.C.L. carried out the calculation and drafted the article with the assistance of Y.Y.L., J.L., H.B.W. and P.H.W.. All authors contribute to the writing and revision of the manuscript.

\section*{FIGURE LEGENDS}
Figure~1: Evolution of superconducting orders and nematic order in a $1.5\lambda \times 1.5\lambda$ region (see Supplementary videos). $\bold{a.1-7}$ is $s$-wave components ${\left| {{\psi _s}} \right|^2}$, $\bold{b.1-7}$ is $d$-wave components ${\left| {{\psi _d}} \right|^2}$, and $\bold{c.1-7}$ is the nematic order $\phi^2$. The legend shows the magnitudes of the order parameters. Each $\bold{1-7}$ corresponds to $t=0.1,t=0.7,t=2,t=5,t=9,t=30$, respectively.\par
\vspace{10pt}

\noindent Figure~2: The vortex lattice (a) is $s$-wave component ${\left| {{\psi _s}} \right|^2}$, (b) is $d$-wave component ${\left| {{\psi _d}} \right|^2}$ and (c) is the nematic order $\phi^2$. (d), (e), (f) are the vortices in FeSe sample under different applied magnetic fields in experiment (Adapted from Ref.~\cite{song2011direct}). The ratio $r$ is defined as the separation of two vortices along the $x$ direction divided by that along the $y$ direction.\par
\vspace{10pt}

\noindent Figure~3: The coordinate is elongated along $y$ direction under the coordinate transformation.\par
\vspace{10pt}

\noindent Figure~4: Nematic order will turn on the $\lambda_1$ interaction to enhance the anisotropy of the vortices. The profile of vortex \textbf{a} without and \textbf{b} turn on the $\lambda_1$ interaction.\par
\vspace{10pt}

\noindent Figure~5: \textbf{a} and \textbf{b} are the profiles of superconducting order ($\psi_s$) with and without nematic order. \par
\vspace{10pt}

\noindent Figure~6: The crystal structure of FeSe, where the $x-$ and $y$-directions are defined as along the Fe-Fe bonds. \par

\begin{figure*}[htbp]
\centering
\includegraphics[width=0.95\textwidth]{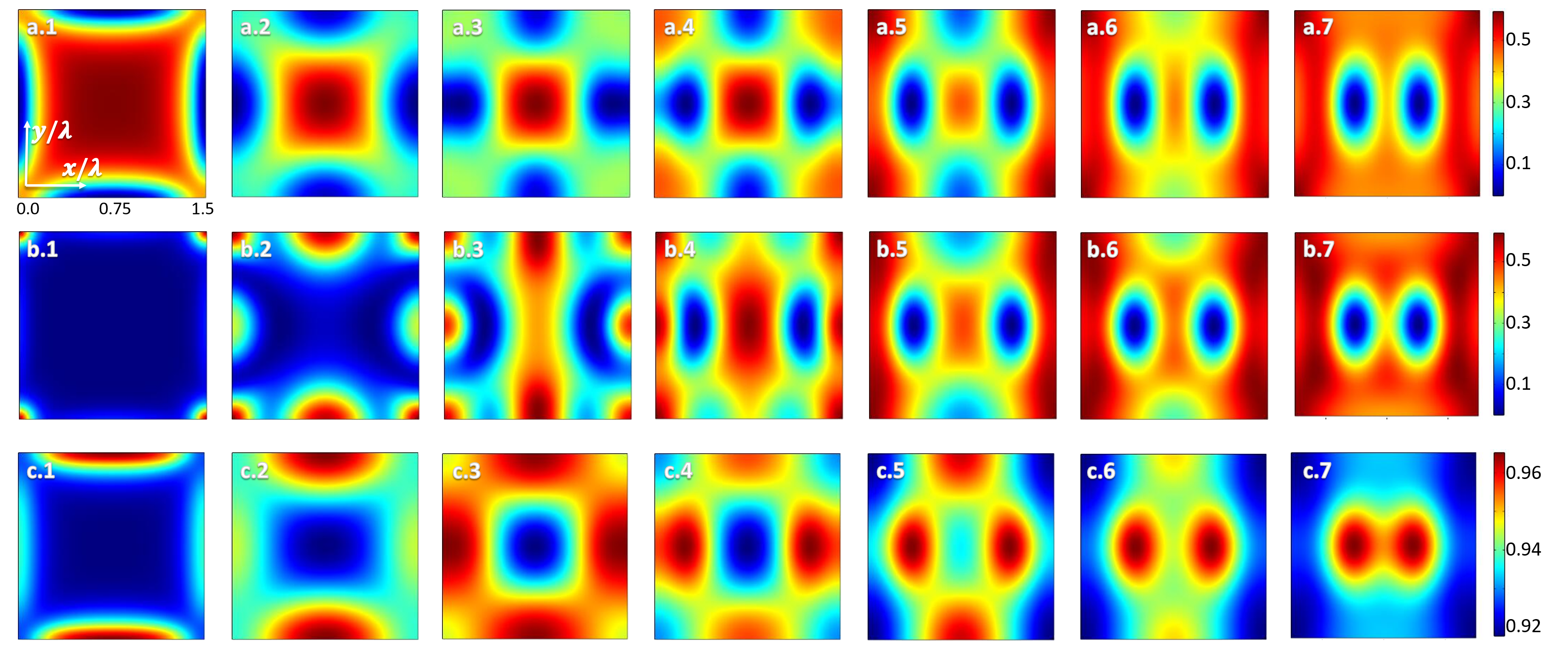}
\end{figure*}
\begin{figure*}[htbp]
\centering
\includegraphics[width=0.95\textwidth]{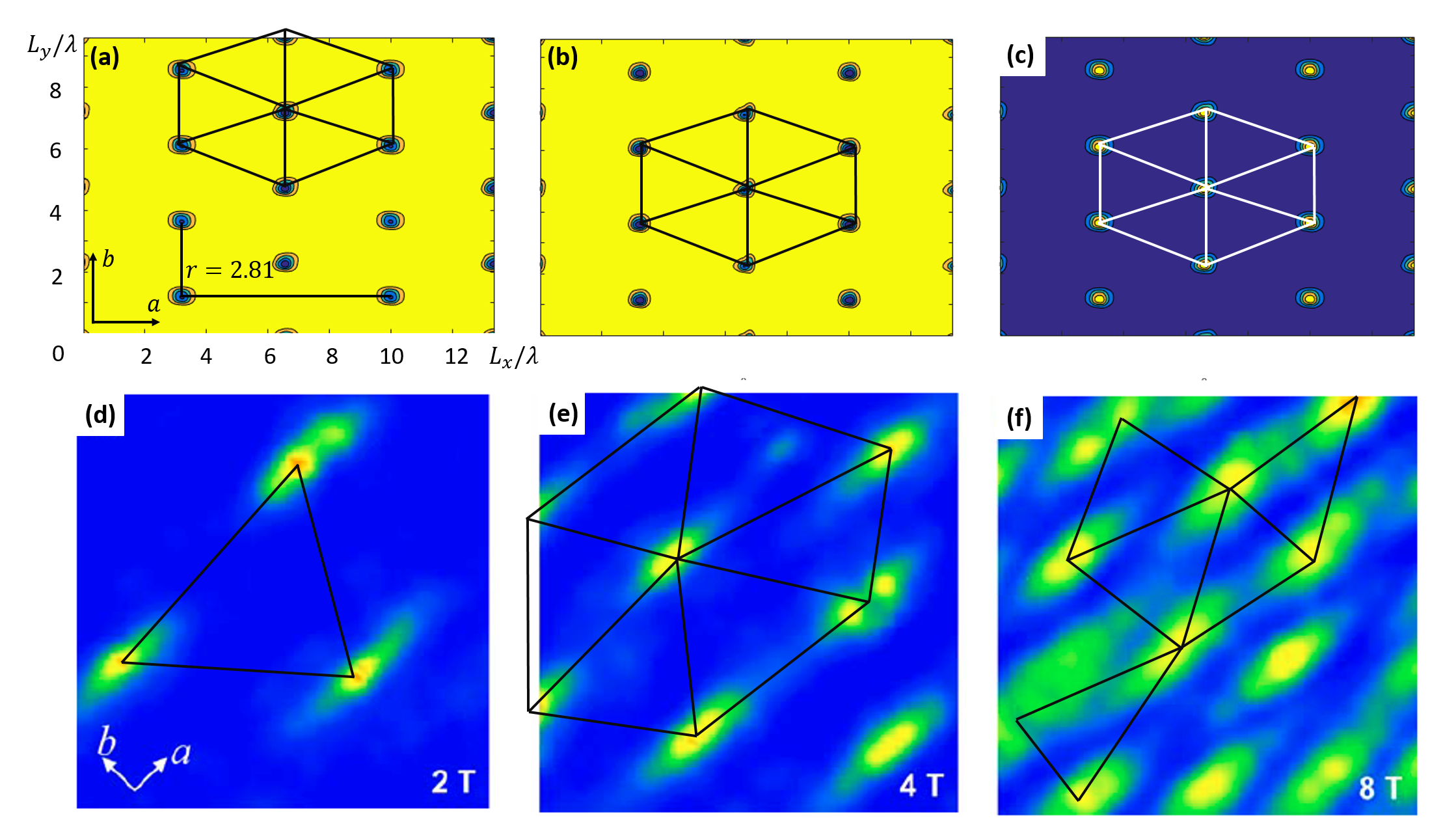}
\end{figure*}
\begin{figure*}[htbp]
\centering
\includegraphics[width=0.45\textwidth]{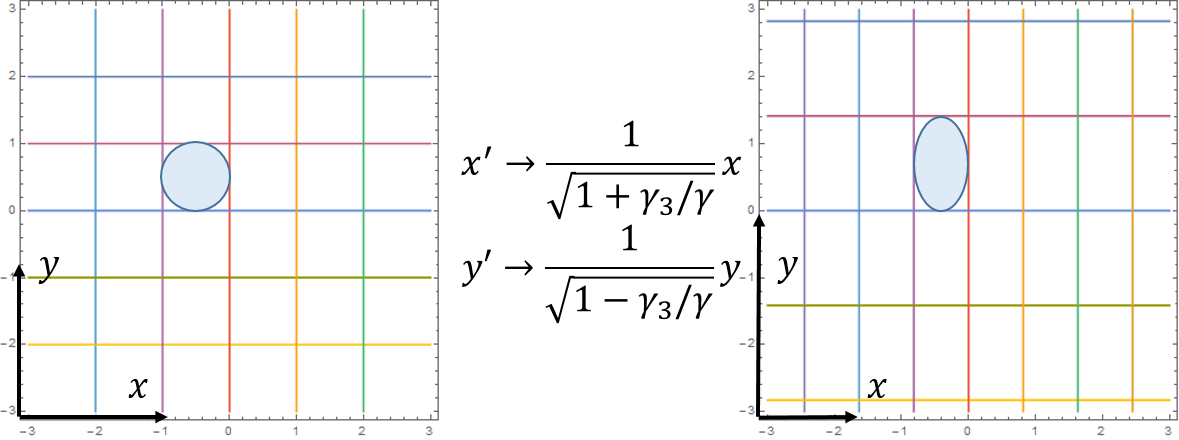}
\end{figure*}
\begin{figure*}[htbp]
\centering
\includegraphics[width=0.45\textwidth]{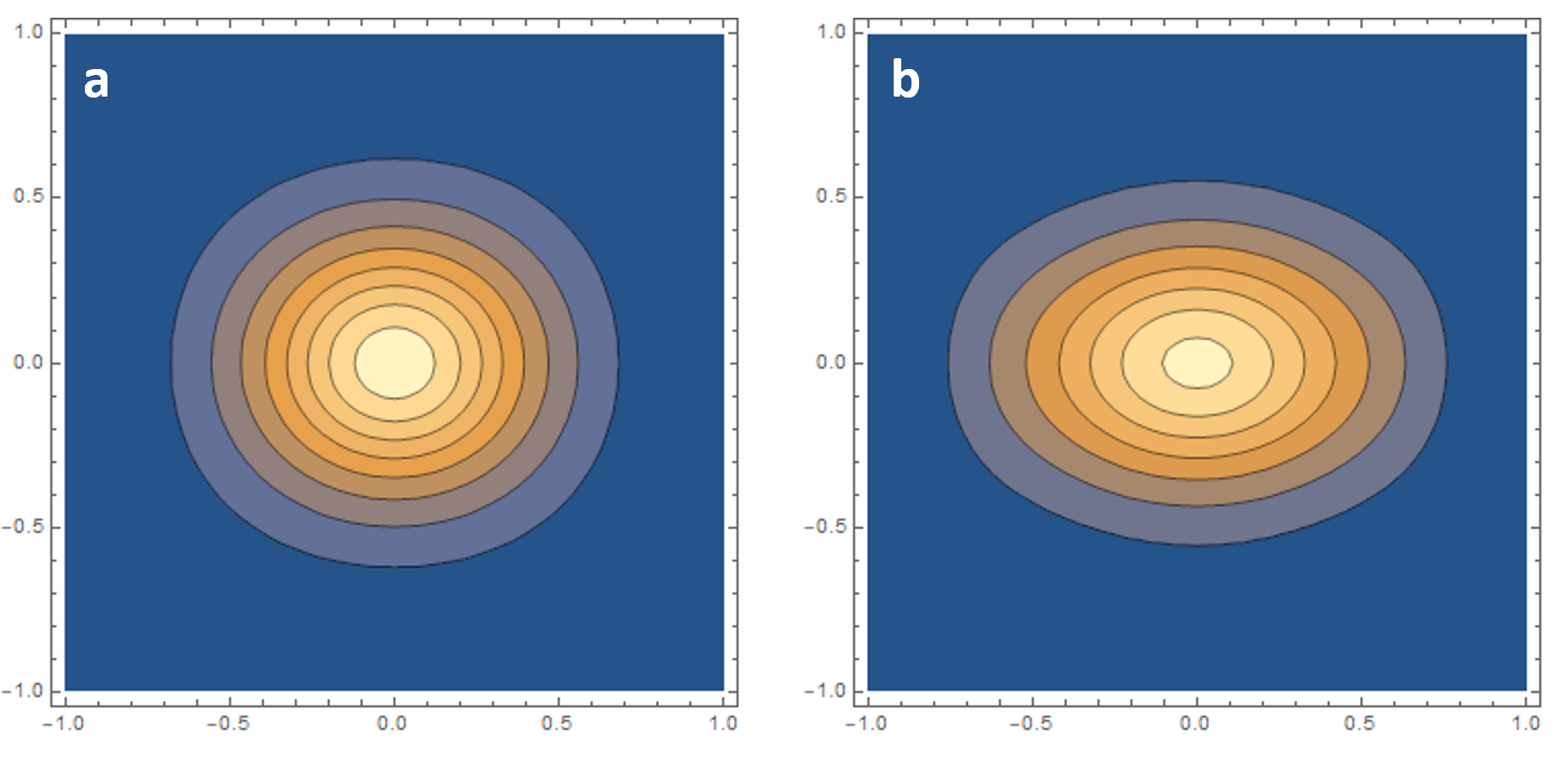}
\end{figure*}
\begin{figure*}[htbp]
\centering
\includegraphics[width=0.45\textwidth]{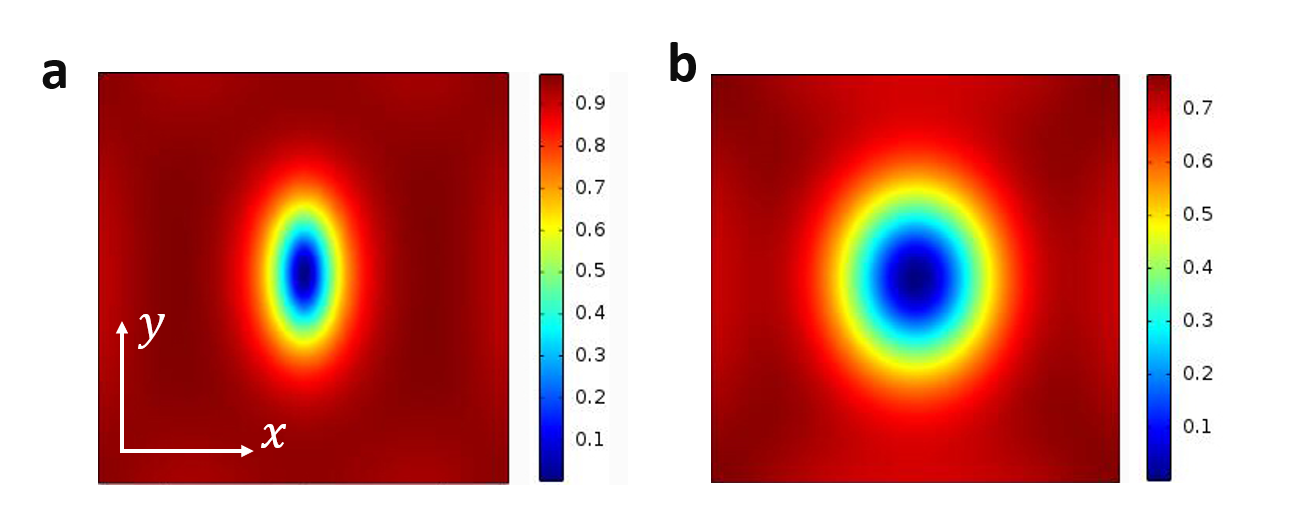}
\end{figure*}
\begin{figure*}[htbp]
\centering
\includegraphics[width=0.45\textwidth]{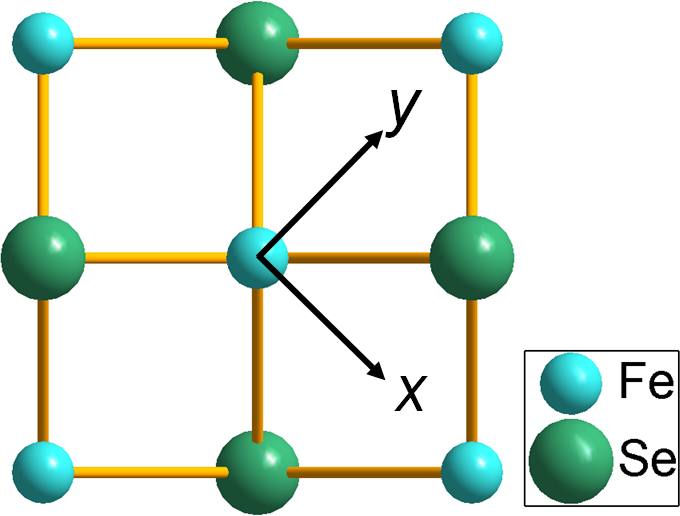}
\end{figure*}

%
%

\end{document}